\documentclass[aip,graphicx, jap, amsmath,amssymb, preprint]{revtex4-1}

\usepackage[english]{babel}
\usepackage[utf8]{inputenc}
\usepackage{amsmath}
\usepackage[colorinlistoftodos, color=green!40, prependcaption]{todonotes}
\usepackage[pdftex, pdftitle={Article}, pdfauthor={Author}]{hyperref} 
\usepackage{graphicx}

\draft 

\begin{document}

\title{Breathing Modes of Skyrmion Strings in a Synthetic Antiferromagnet Multilayer} 

\author{Christopher E. A. Barker}
\email[Correspondence email address: ]{pyceab@leeds.ac.uk}
\affiliation{School of Physics and Astronomy, University of Leeds, Leeds LS2 9JT, United Kingdom}
\affiliation{Bragg Centre for Materials Research, University of Leeds, Leeds LS2 9JT, United Kingdom}

\author{E. Haltz}
\altaffiliation{Present address: Universit\'{e} Sorbonne Paris Nord, 99 Av. Jean Baptiste Cl\'{e}ment, 93430 Villetaneuse, France}
\affiliation{School of Physics and Astronomy, University of Leeds, Leeds LS2 9JT, United Kingdom}

\author{T.A. Moore}
\affiliation{School of Physics and Astronomy, University of Leeds, Leeds LS2 9JT, United Kingdom}

\author{C.H. Marrows}
\email[Correspondence email address: ]{c.h.marrows@leeds.ac.uk}
\affiliation{School of Physics and Astronomy, University of Leeds, Leeds LS2 9JT, United Kingdom}
\affiliation{Bragg Centre for Materials Research, University of Leeds, Leeds LS2 9JT, United Kingdom}

\date{\today}

\begin{abstract}
Skyrmions are small topologically protected magnetic structures that hold promise for applications from data storage to neuromorphic computing and they have been shown to possess internal microwave frequency excitations. Skyrmions in a synthetic antiferromagnet have been predicted to be smaller and faster than their ferromagnetic equivalents, and also shown to possess more internal modes. In this work we consider the breathing modes of skyrmions in a four repetition synthetic antiferromagnetic multilayer by means of micromagnetic simulations and examine the further splitting of the modes into different arrangements of out-of-phase, in-phase and modes with more complex phase relationships. This results in a lowering of frequencies which is promising for skyrmion sensing applications in a synthetic antiferromagnet. 
\end{abstract}

\maketitle 

\section{Introduction}
	
Magnetic skyrmions are topologically protected magnetisation textures where the core spin orients antiparallel to the surrounding magnetisation~\cite{Nagaosa2013, Everschor-Sitte2018}. They are stabilised in bulk materials by the Dzyaloshinkii-Moriya interaction (DMI) at a narrow range of temperatures just below the ferromagnetic ordering temperature~\cite{Muhlbauer2009,Wilhelm2011, Adams2012}. In magnetic thin films they can be stabilised at room temperature by a competition between the perpendicular magnetic anisotropy and the DMI across an interface between a magnetic layer and a non-magnetic heavy metal layer~\cite{Woo2016, Jiang2017}. It has been proposed that they could be the `bits' in a new generation of magnetic storage devices~\cite{Tomasello2014}, and their fundamental properties are of great interest~\cite{Buttner2018}. There remain, however, some key challenges in applying skyrmions to real devices. Reducing their size, enhancing their robustness at room temperature, and lowering the power required to manipulate them are all active areas of research~\cite{Back2020, Vakili2021}.\\\\

In order to overcome these challenges it has been proposed to use synthetic antiferromagnets (SAFs)~\cite{Duine2018}. In such a system, two ferromagnetic layers are coupled antiferromagnetically through a non-magnetic spacer layer in an RKKY-style interaction. Skyrmions in these systems have been predicted to be smaller, more stable, and require less energy to manipulate~\cite{Zhang2016, Zhang2016_PRB}. Recent experiments have also shown that skyrmions can be stabilised~\cite{Legrand2019} in such materials, and---in the case of large skyrmion bubbles---driven by applied current pulses~\cite{Dohi2019}. However new challenges are introduced. Because of the cancellation of the stray field imaging or sensing skyrmions in these systems is challenging. 
\\\\

Skyrmions also possess dynamic modes in the microwave frequency range. They either gyrate (rotational modes) or coherently expand and contract (breathing modes) about their centre of mass. These modes are well understood both theoretically~\cite{Mochizuki2012} and experimentally~\cite{Onose2012} in bulk systems, and are used routinely to characterise the extent of the skyrmion pocket in phase space~\cite{Birch2019}. In thin film systems this is much more difficult because of the lowered sensitivity of measurements and the reduced number of skyrmions in the system~\cite{Satywali2021}, and so most work on these systems is theoretical~\cite{Kim2014, Lonsky2020_APLMaterials}. It was recently shown that in synthetic antiferromagnets, the coupling between the skyrmions in each layer produces a splitting of both the rotational~\cite{Xing2018} and breathing~\cite{Lonsky2020_PRB} modes into in-phase and out-of-phase modes. This is of interest because in the out-of-phase mode one skyrmion is at its maximum radius while the other is at its minimum, so a measurable trace can be observed in the GHz response of the system, unlike for the in-phase mode. On this basis it has been proposed that a broadband FMR would be a robust tool to detect and identify chiral magnetic textures in a SAF~\cite{Lonsky_2022}.\\\\

In this work, we study the breathing modes of skyrmions in SAFs using the micromagnetic solver \textsc{mumax3}~\cite{Vansteenkiste2014}, and expand on previous work to consider SAFs with multiple repeats. In such samples we see further splitting of modes, revealing a variety of complex phase relationships and yielding resonances at lower frequencies more accessible by conventional laboratory equipment, paving the way to experimental observation of such modes.

\section{Methods}
	
\begin{figure}
\includegraphics[width=\linewidth]{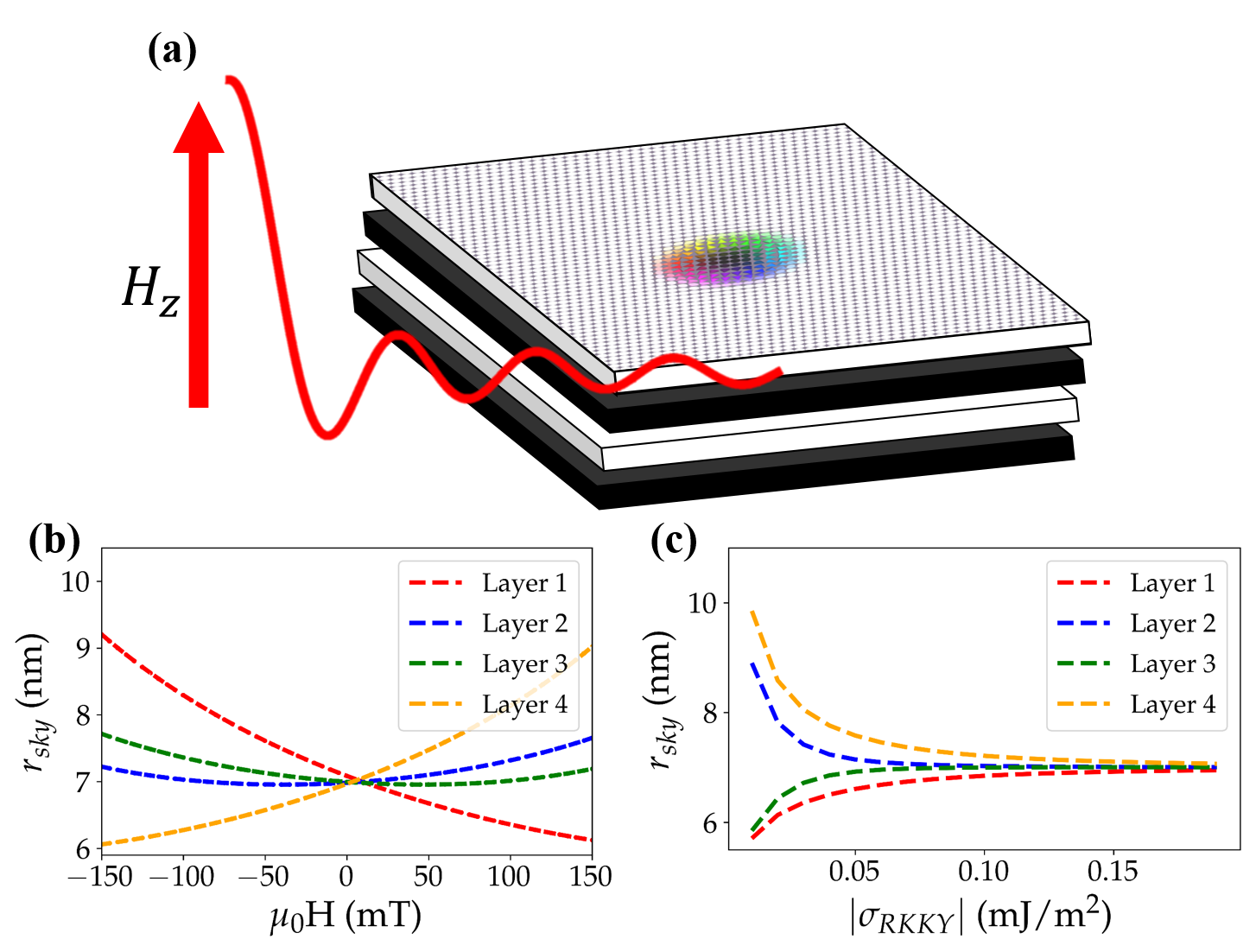}
\caption{(a). Schematic of magnetic system considered along with illustration of the out of plane sinc excitation pulse. (b,c). Radius of the skyrmions in each layer as a function of; (b) applied magnetic field at an RKKY strength of 0.2~mJ/m$^2$, and (c) interlayer RKKY coupling strength in an applied field of 50~mT.}
\label{fig_1} 
\end{figure}

In this study we model our system using the finite difference micromagnetic solver \textsc{mumax3}. This numerically solves the Landau-Lifshitz-Gilbert (LLG) given by
\begin{equation}
    \dfrac{d\mathbf{m}}{dt} = -\lvert \gamma_0 \rvert \mathbf{H}_{\mathrm{eff}} \times \mathbf{m} + \alpha \mathbf{m} \times \dfrac{d\mathbf{m}}{dt} .
    \label{LLG}
\end{equation}
Here $\mathbf{m}$ is the unit vector of the magnetisation, $\gamma_0$ is the gyromagnetic constant, and $\mathbf{H}_{eff}$ is the effective magnetic field
\begin{equation}
    \mathbf{H}_{\mathrm{eff}} = -\dfrac{1}{\mu_{\mathrm{0}} M_{\mathrm{s}}}\dfrac{\delta U}{\delta \mathbf{m}},
    \label{H_eff}
\end{equation}
which is proportional to the variation in the total micromagnetic energy $U$. $M_s$ is the total saturation magnetisation, and $\mu_0$ is the permeability of free space. $U$ contains terms describing the Zeeman energy, the isotropic exchange interaction, demagnetisation effects, a uniaxial anisotropy perpendicular to the plane of the film and an interfacial Dzyaloshinkii-Moriya interaction (DMI), whose energy density is given by

\begin{equation}
    U_{\mathrm{DMI}} = D [ m_z(\nabla \cdot \mathbf{m} ) - (\mathbf{m}\cdot \nabla)m_z ],
    \label{DMI}
\end{equation}
where D is the DMI constant and controls the strength of the interaction. The system is composed of four magnetic layers of thickness 1~nm separated by nonmagnetic spacer layers also of thickness 1~nm. Each magnetic layer is coupled antiferromagnetically to both the one above and below it in such a way that in the ground state of the system the magnetisation of the layers alternates up the stack. A schematic of the stack is shown in figure \ref{fig_1}a. Each layer is a 100~nm~$\times $~100~nm square composed of 64 cells such that the $x$-$y$ cell size is 1.5625~nm, suitably smaller than the exchange length for this system $l_0 = 4.34$~nm. The magnetic parameters of the system are defined explicitly within the simulation, with the saturation magnetisation $M_{\mathrm{s}}=0.796\times 10^6$~A/m, exchange stiffness $A = 15\times 10^{-12}$~J/m, uniaxial anisotropy $K = 1\times 10^6~\mathrm{J/m^3}$, interfacial Dzyaloshinskii-Moriya interaction $D = 3\times 10^{-3}~\mathrm{J/m^2}$ and Gilbert damping parameter set to $\alpha = 0.01$ in order to resolve the high-frequency dynamics of the system. These values were chosen to represent typical material parameters of systems shown to host skyrmions in the literature~\cite{Jiang2017}. Periodic boundary conditions were applied in order to eliminate any edge effects from the borders of the simulation.\\\\
The simulation was prepared with a N\'eel skyrmion initialised in the centre of each layer of alternating core orientation and chirality. The entire system was then allowed to relax to its energy minimum with (unless otherwise specified) a static magnetic field $H_0= 50$~mT applied along the positive $z$ direction. Once the skyrmions reach their equilibrium size (shown in figure 1(b,c)), an oscillating $H_z$ field was applied starting at $t = 0$ of the form
\begin{equation}
    H_z = H_0 + \dfrac{H_1 \sin(2\pi f_{\mathrm{MAX}} t)}{2\pi f_{\mathrm{MAX}} t} ,
    \label{excitation_field}
\end{equation}
where $H_1$ the amplitude of the excitation field is set to 50~mT and $f_{\mathrm{MAX}}$ the cut-off frequency of the excitation field is 100~GHz. This function was chosen as it is a square in the frequency domain and so excites all frequencies up to the cut-off frequency equally. After this excitation pulse the system was allowed to run for 20~ns with the total magnetisation of the system as well as the magnetisation of each individual layer recorded every 2~ps. In order to calculate the frequency response of the system we use the power spectral density (PSD), which is calculated using equation \ref{PSD} and is the Fourier transform of the variation of the magnetisation multiplied by its complex conjugate:
\begin{equation}
    \mathrm{PSD} = \Big\vert \int_{0}^{t_0} dt \exp ( 2 \pi f t ) \delta m_z(t) \Big\vert ^2 .
    \label{PSD}
\end{equation}
The variation of the magnetisation as a function of time in this expression is given by $\delta m_z (t) = \langle m_z (0) \rangle - \langle m_z(t) \rangle$, which is the difference between the spatial average of the magnetisation at $t=0$ and a given time $t$.

\section{Results and Discussion}

\subsection{Static Properties of Skyrmions in a SAF}

\begin{figure} 
    \includegraphics[width=\linewidth]{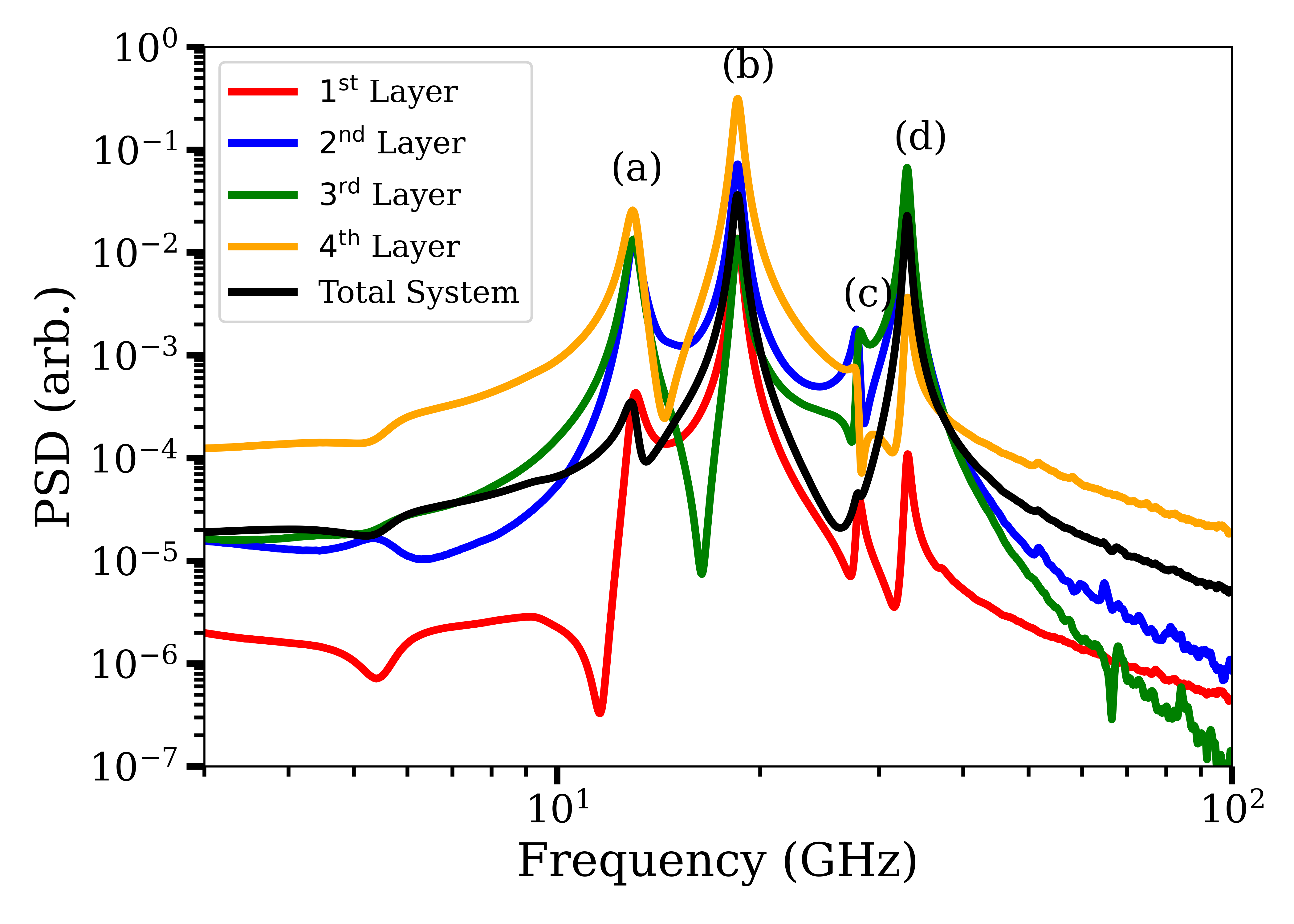}
    \caption{Power Spectral Density of SAF multilayer system after application of sinc field pulse at $t=0$ in a static field of 50~mT. The frequency response of the total system is plotted in black, along with the individual layers in colour.}
    \label{fig_2}
\end{figure}

\begin{figure*}
    \includegraphics[width = \linewidth]{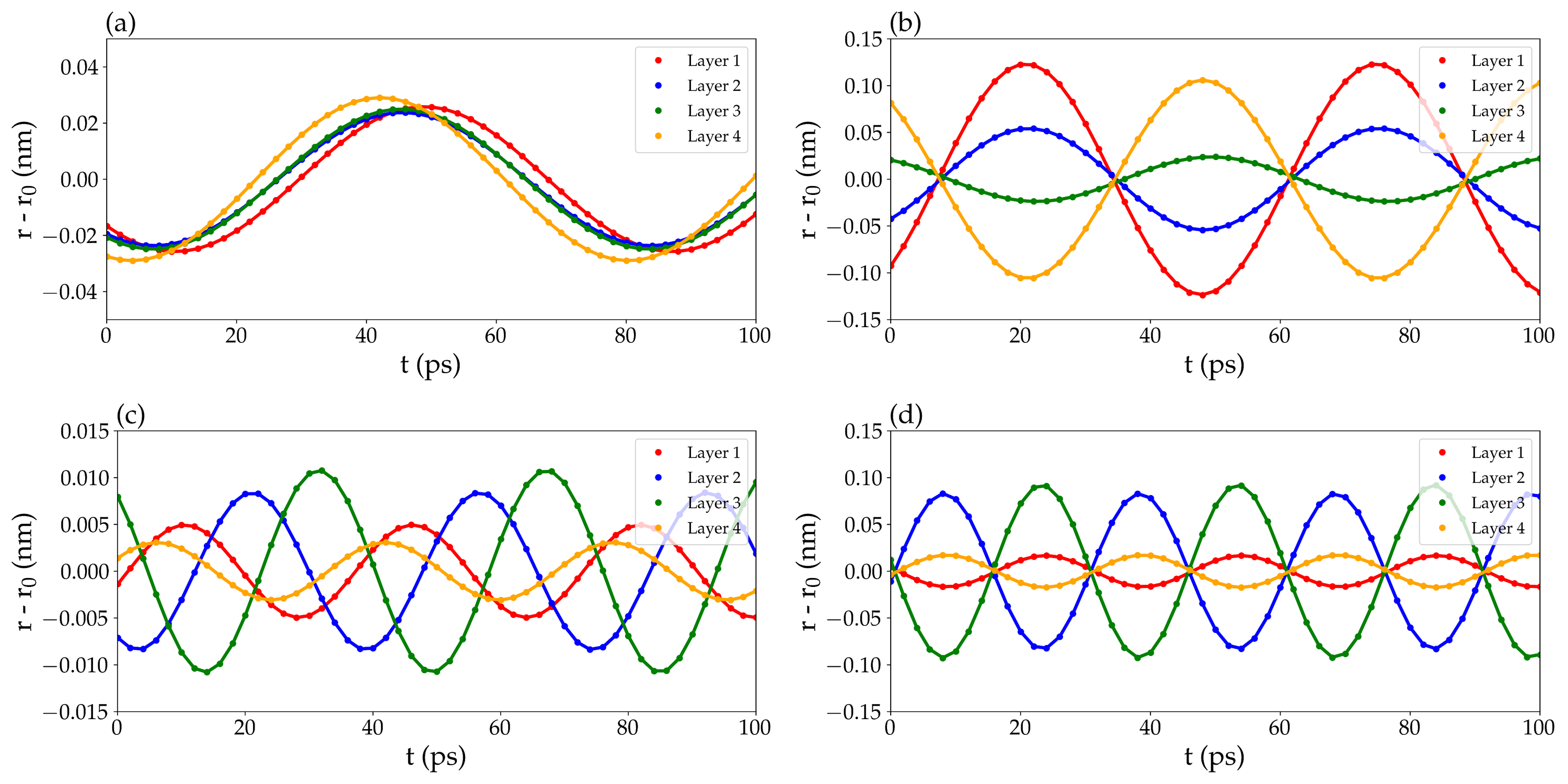}
    \caption{(a-d). Real space variation of skyrmion radius in each layer as a function of time driven at the frequencies of each of the peaks denoted in figure \ref{fig_2}.}
    \label{fig_3}
\end{figure*}

In order to support our understanding of the variation of breathing modes of skyrmions in a SAF we first consider their static properties. Figures \ref{fig_1}(b,c) show the variation in the radius of the skyrmions as a function of applied magnetic field (b) and strength of the RKKY interaction (c). Skyrmion radius is calculated as half the distance between zero crossings of the $z$ component of the magnetisation. As the magnetic field increases, the size of the skyrmions in the two layers whose cores align with the field also increases, as we might expect. However as we see in figure \ref{fig_1}b, while the size of the outermost two skyrmions initially decrease, the two in the inner layers reach a minimum before increasing in size again. This is a result of being sandwiched between two increasing skyrmions and the desire of the system to exist in a state where spins in alternating layers are equal and opposite across all space, which cannot be satisfied if the skyrmion were to simply decrease in radius. Thus the competition between the RKKY interaction and the Zeeman energy results in the skyrmion also increasing in size. In the case of the skyrmion in the outer layer, this competition also takes place, but is considerably weaker as it is only affected by one nearest neighbour, so it continues to decrease in size, albeit at a decreasing rate as the RKKY energy barrier increases with increasing disparity in skyrmion radius. The radius of the skyrmions as a function of the antiferromagnetic coupling strength was also measured in a 50~mT static biasing field along the $z$ axis. For suitably weak RKKY strengths, $\sigma_{\mathrm{RKKY}}$, the radii of the skyrmions diverges due to the Zeeman interaction, however as the strength of the RKKY increases the radii converge to the same size which remains constant for all larger $\sigma_{\mathrm{RKKY}}$. For all further results, unless otherwise specified an RKKY coupling strength of $\sigma_{\mathrm{RKKY}}$~=~-0.2~mJ/m$^2$ was chosen, thus in the static condition the radii of all skyrmions are equal.

\subsection{Dynamics}

\begin{table}
    \caption{Relative phase in degrees of the breathing of the skyrmions in each layer when driven by a sinusoidal driving field at frequencies denoted in figure~\ref{fig_2}. Phases have been shifted so the values presented are always the phase shift from layer 1.}
    \begin{ruledtabular}
    \begin{tabular}{l c c c c}
        & peak (a) & peak (b) & peak (c) & peak (d)\\
        \hline
        Layer 1 &0$^\circ$ &0$^\circ$ &0$^\circ$ &0$^\circ$\\
        Layer 2 &15.2(1)$^\circ$ &5.2(1)$^\circ$ &106.9(1)$^\circ$ &173.6(1)$^\circ$\\
        Layer 3 &16.0(1)$^\circ$ &190.8(1)$^\circ$ &211.8(1)$^\circ$ & 6.9(1)$^\circ$\\
        Layer 4 &31.2(1)$^\circ$ &181.6(1)$^\circ$ &317.9(1)$^\circ$ &179.0(1)$^\circ$ \\
        
    \end{tabular}
    \end{ruledtabular}
    \label{phi_table}
    
\end{table}

\begin{figure}[t!]
    \includegraphics[width=\linewidth]{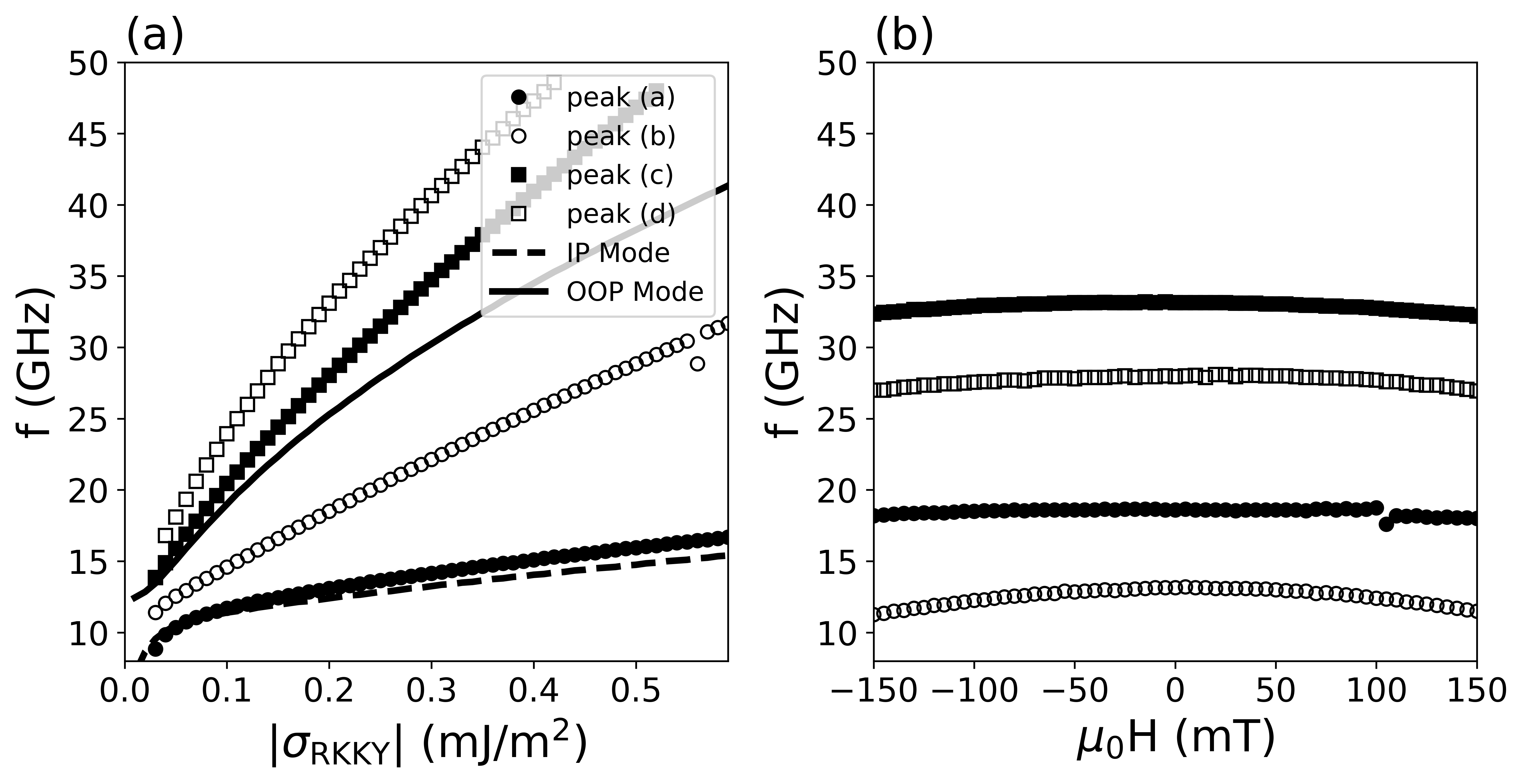}
    \caption{(a). Peak frequencies of the SAF multilayer after application of a sinc field pulse starting at $t=0$ in a static field of 50~mT as a function of the antiferromagnetic coupling strength. Points show the frequencies of the peaks corresponding to panels (a-d) in figure 3, while the lines correspond to the in-phase (IP) and out-of-phase (OOP) modes for a bilayer SAF. (b) Peak Frequencies as a function of applied magnetic field.}
    \label{fig_4}
\end{figure}

The power spectral density of the system is shown in figure~\ref{fig_2} when excited by a sinc pulse in a $H_0=$~50~mT biasing field. The black curve shows the frequency response of the total system, whereas those of the individual layers are shown in the other colours. Peaks are labelled a,b,c,d to correspond to the same panels of figure~\ref{fig_3}. At peak (a), a strong response is observed in each of the individual layers, whereas the response of the total system is several orders of magnitude lower at this frequency. This suggests that at this peak, each skyrmion is breathing in phase with all of the others, thus over all space the net magnetisation of the system remains constant and so there is very little trace of the breathing mode. The small peak can be explained by the presence of the static biasing field inducing small differences in the sizes of the skyrmions and thus allowing us to see a (albeit small) trace of the breathing mode. On the other hand, at peak (b) we see a strong response in the total system as well as in each of the individual layers, suggesting an inverse relationship between the breathing modes of each layer, meaning the system oscillates between a net positive and net negative moment and there is a detectable signal. Similarly for the two higher frequency peaks, (c) and (d), there is almost no discernible trace of the first peak in the response of the total system, whereas there is a very strong signal for the second.\\\\

This interpretation is confirmed in figure~\ref{fig_3}, which shows the variation in the radius of the skyrmions in each layer when driven by a sinusoidal driving field at the frequency of each of the peaks in figure~\ref{fig_2} and capturing the magnetisation at intervals of 2~ps. Panels (a-d) correspond to the peaks labelled (a-d) in figure~\ref{fig_2}. Points correspond to the measured radius of the skyrmions, whereas solid lines are fits to a general sinusoidal function of the form $F(t) = I\sin (2\pi f t - \phi)$. Values of the phase corresponding to all of these fits are shown in Table~\ref{phi_table} which aid in our interpretation of panel (c) in particular. There we see that the phase varies by approximately $\pi$/4 between each layer, in such a way that if we consider a vertical slice through the system, the total magnetisation remains essentially unchanged, hence resulting in there being no features present in the PSD of the entire system.\\\\

This more complex behaviour compared to the simple system presented in~\cite{Lonsky2020_PRB} can be explained by considering the symmetry breaking of the system~\cite{Kim2018}. Each additional skyrmion in the system increases the number of possible arrangements of the breathing modes, as would be expected for any system of coupled harmonic oscillators. The ordering of the mode frequency is in order of energy minimisation, where in our systems the dominant interaction is the RKKY-style antiferromagnetic coupling. Thus the lowest energy mode is the one where the skyrmions breathe in phase with the same amplitude, as thus the total magnetisation through the system remains constant at all time $t$ and the RKKY energy is minimised. The frequency splitting between the two out-of-phase modes can be understood in a similar fashion. Because of the antiferromagnetic coupling between layers, each layer is effectively screened from all but its nearest neighbours. The mode in figure 3(b) is only out of phase between the two central layers, while in phase between the two outer pairs of layers. Thus the net breaking of the antiferromagnetic coupling is considerably less than for the mode in figure 3d where each skyrmion breathes out-of-phase with both the one above and below it.\\\\

This also explains the behaviour shown in figure \ref{fig_4}a, which shows the frequency of the breathing modes as a function of the antiferromagnetic coupling strength $\sigma_{\mathrm{RKKY}}$. Their frequency increases with $\sigma_{\mathrm{RKKY}}$ as we might expect as the energy barrier due to the antiferromagnetic coupling to overcome is larger. The dashed and solid lines show the corresponding frequencies of the in-phase and out-of-phase peaks found when considering a simple SAF bilayer. As we can see the frequency of the lowest peak matches that of the bilayer skyrmion, while the two highest peaks are at a higher frequency than the out-of-phase peak. This leaves the first out-of-phase peak of our SAF multilayer, which lies at a considerably lower frequency than the equivalent in the SAF bilayer. Thus by extending the number of layers we are able to lower the frequency of the first observable mode as multiple arrangements become possible, including ones with lower energy than the single out-of-phase mode found in the bilayer system. We would expect this effect to scale with the number of layer repetitions, and thus make synthetic antiferromagnetic multilayers with high numbers of repetitions the ideal media in which to observe these low frequency out-of-phase breathing modes.\\\\

Figure~\ref{fig_4}b shows the variation of the breathing mode frequency with magnetic field, which varies minimally over the field range considered, despite the strong variation in skyrmion radius in figure~\ref{fig_1}b. This is in contrast to the behaviour in ferromagnetic systems~\cite{Kim2014} where the frequency of the breathing mode was shown to vary with field. The only features evident at higher fields are the appearence of a trace of the in-phase breathing modes. This is as a result of the Zeeman effect discussed in section IIIA, meaning that the disparity between the initial sizes of the skyrmions is such that a measurable frequency response can be observed. This goes to demonstrate the strong effect of the antiferromagnetic coupling on the system, dominating all other interactions. Similarly, while changing other magnetic parameters shifts the frequencies of the modes, the relation between them remains the same.
\subsection{Effects of Structural Disorder}
	
\begin{figure}
    \includegraphics[width=\linewidth]{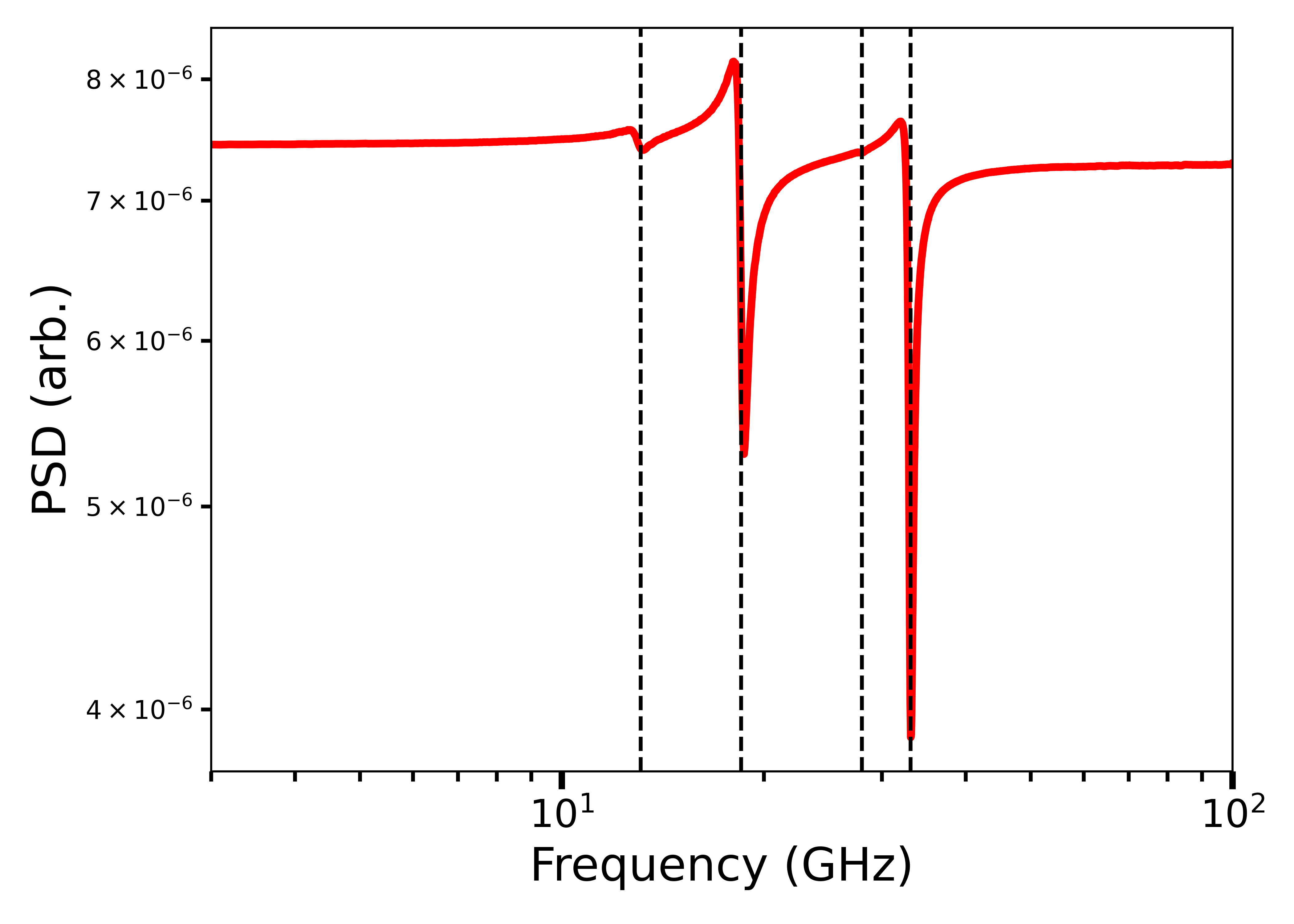}
    \caption{Total frequency response of a 4-magnetic layer SAF multilayer with disorder introduced in the $M_s$. Dotted lines show expected mode frequencies from figure~\ref{fig_2}}
    \label{disordered_dynamics}
\end{figure}

While these results are fascinating as a way forward to observing the GHz response of skyrmions in a synthetic antiferromagnet, these simulations are all performed in perfectly smooth layers at zero temperature. In order to gain an insight on the effect of disorder in our systems we performed a set of simulations with random variations in the saturation magnetisation.\\\\
To do this we allow the $M_s$ of each lateral cell in our simulation to vary with a normally distributed value capped at $\pm$~10\% of the global $M_s$ of the system. After initialising the system with a skyrmion in the centre we allow it to relax to its equilibrium size. Then we introduce the random variations in the $M_s$ and allow the system to relax again. The skyrmion moves slightly to a local area of reduced free energy and otherwise does not change its size, shape or antiferromagnetic coupling.
We can then excite the disordered system with the same sinc pulse as before, and use a Fourier transform to examine the frequency response of the system.\\\\
The results of comparing the disordered system to the measurements in the main manuscript are presented in figure~\ref{disordered_dynamics}. The frequency response of the total system is plotted in red, along with the frequencies of each peak from figure~\ref{fig_2} plotted as black dashed lines. We see no change in frequency of each peak; simply a damping in the amplitudes of the peaks. The behaviour of the system - arranging itself into out-of-phase detectable modes and in-phase `hidden' modes remains the same, suggesting that such behaviour would not fundamentally change in a real system.

\subsection{Effect of a Finite Temperature}

\begin{figure}
    \includegraphics[width=\linewidth]{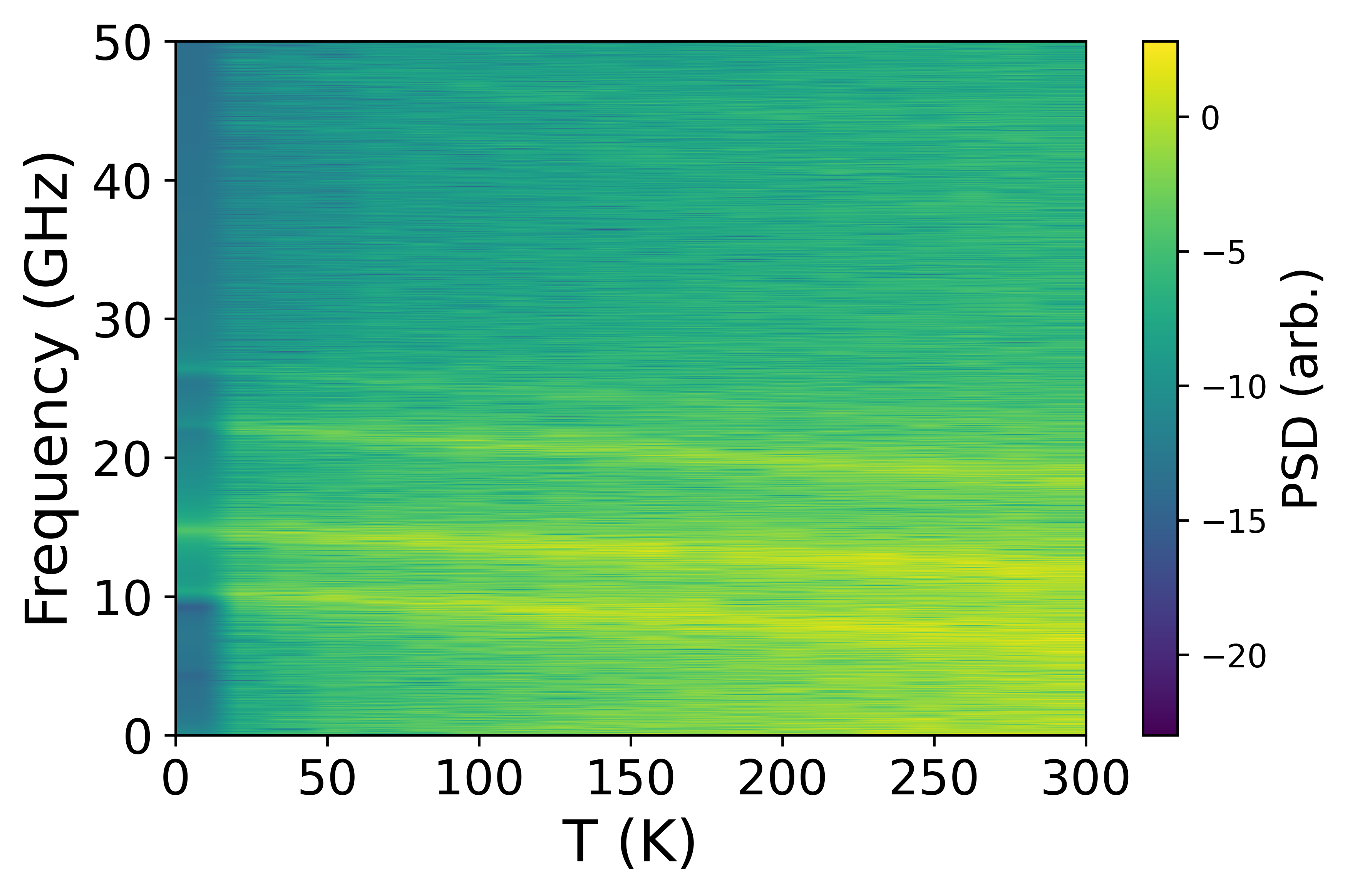}
    \caption{Frequency of breathing modes vs temperature. Magnetisation dataset taken from the bottom magnetic layer in order to resolve all modes.}
    \label{fvT}
\end{figure}

In order to further examine the effects of true experimental conditions we use the thermal field feature of \textsc{mumax3}~\cite{Vansteenkiste2014} to simulate the frequency response spectrum at a range of temperatures between 0~K and 300~K. The results of this are shown in figure~\ref{fvT}. We see an expected decrease that appears to be close to linear in the frequency of each of the four modes with temperature, and an increase in the linewidth of the peaks as well as the noise levels in the spectrum. However the fundamental frequency relationship between modes remains unchanged, as does the number of total modes.

\section{Conclusion}
	
In this work we simulated a synthetic antiferromagnetic multilayer composed of a total of four magnetic layers, each one coupled antiferromagnetically to the one above and below it. We excited the breathing modes of the system by applying an out of plane sinc field pulse and measuring the time dependence of the total magnetisation. In doing so we find that the two modes observed in single SAFs~\cite{Lonsky2020_PRB} split further into four modes. These arrange themselves into two in-phase breathing modes, and two out-of-phase breathing modes.\\
It has been proposed that measurement of the microwave frequency response of skyrmions in a SAF could present a new method of skyrmion detection in an otherwise challenging system due to the negligible stray field. Our results show that in a SAF multilayer the frequency of the first observable mode is considerably lowered in comparison with the case of a bilayer SAF. This is advantageous for observation with typically available microwave measurement apparatus, and the trend of lowered frequencies as a function of layer repetitions makes SAF multilayers with high numbers of repetitions the ideal candidate in which to observe these modes.\\ In addition, we are able to simulate the effects of structural and thermal disorder on our system, and show that they do not have a demonstrable effect on the fundamental physics presented in the main body of our work. Realising these measurements in experimental conditions is difficult and has only recently been shown~\cite{Satywali2021}, however we believe our results show that it is possible in a synthetic antiferromagnetic multilayer. 

\begin{acknowledgments}
The authors would like to thank J. Barker and G. Burnell for useful discussions about the work. This work was supported by EPSRC grant number EP/T006803/1. C.E.A. Barker acknowledges funding from the National Physical Laboratory. 
\end{acknowledgments}

\bibliography{aipsamp}

\end{document}